\documentclass{acm_proc_article-sp}
\usepackage{url}
\usepackage{enumitem}
\usepackage{amssymb,amsmath}
\begin{document}
\setenumerate{itemsep=2pt,topsep=0pt,parsep=0pt,partopsep=0pt,leftmargin=1.5pc}
\def\etal{\emph{et~al.}\,}
\def\ie{\emph{i.e.}\,}
\newcommand{\fullset}[2]{\{#1_1,\dots,#1_#2\}}


\title{Indicators of Good Student Performance in \\Moodle Activity Data}
\numberofauthors{3}
\author{
\alignauthor
Ewa M\l{}ynarska\\
       \affaddr{Insight Centre, University College Dublin, Ireland}\\
       \email{ewa.mlynarska@insight-centre.org}
\alignauthor
Derek Greene\\
       \affaddr{Insight Centre, University College Dublin, Ireland}\\
       \email{derek.greene@ucd.ie}
\and        
\alignauthor P\'{a}draig Cunningham\\
       \affaddr{Insight Centre, University College Dublin, Ireland}\\
       \email{padraig.cunningham@ucd.ie}
}
\maketitle


\begin{abstract}
In this paper we conduct an analysis of Moodle activity data focused on identifying early predictors of good student performance. The analysis shows that three relevant hypotheses are largely supported by the data. These hypotheses are: early submission is a good sign, a high level of activity is predictive of good results and evening activity is even better than daytime activity. 
We highlight some pathological examples where high levels of activity correlates with bad results. 

\end{abstract}


\section{Introduction}

The availability of log data from virtual learning environments (VLEs) such as Moodle presents an opportunity to improve learning outcomes and address challenges such as high levels of student dropout \cite{agnihotribuilding,corrigan2014,Siemens2011Penetrating}.
Research has shown that certain activity patterns are indicative of good student performance. At the most basic level, it is typically the case that higher levels of activity correlates with good grades 
\cite{caseymining}. Digging deeper, it has been shown that work submitted close to the deadline is less likely to score well \cite{arnott2014time} and that evening activity is a better predictor of good performance than daytime activity \cite{caseymining}. 

The main contribution of this paper is to analyse a large volume of Moodle activity data to determine whether it can provide any early indicators of good or poor student performance.
We use the data to test three fundamental hypotheses that have been proposed in a number of recent studies:
\begin{enumerate}
\item Submitting an assignment well in advance of the deadline is predictive of students achieving good grade \cite{arnott2014time}.
\item A high level of Moodle activity before submission is predictive of a student achieving good performance \cite{caseymining}.
\item Evening Moodle activity is more predictive of good performance than daytime activity \cite{caseymining}.
\end{enumerate}

We find that these hypotheses are largely borne out in the context of our data (see Section \ref{sec:hypotheses}). However, we do observe some anomalous cases where high levels of activity on certain assignments are negatively correlated with high grades. This indicates situations where a student's effort (in terms of Moodle activity) does not appear to be rewarded with good results. 

In the next section we provide a summary of previous research on educational data mining that is relevant to our work. Section \ref{sec:hypotheses} describes the Moodle data that is analysed in our paper, and reports on the extent to which the data supports the three fundamental hypotheses outlined above. The paper finishes with some conclusions and plans for future work in Section \ref{sec:Conc}.

\section{Related Work}

Initial studies related to the key performance hypotheses described in Section 1 were performed by Casey and Gibson \cite{caseymining}. Two of their hypotheses are partially covered in our work: 1) do Moodle page views correlate with final grades? 2) does it matter if students access Moodle resources on or off campus? The authors find that the higher the level of Moodle activity, the higher the student grades. Activity off campus and in the evenings also showed correlation with higher grades. 
Casey and Gibson \cite{16} examined Moodle log data from Computer Science courses to understand student behaviour, focusing on the relation between resource view counts and final course grades. They concluded that the correlation between Moodle activity and student performance was mostly positive, although the hypothesis did not hold true for Masters courses.

Following work performed in \cite{agnihotribuilding} covering ``at risk'' students,  Corrigan \cite{corrigan2014} continued the idea of applying machine learning models in this area, by using a Support Vector Machine classifier to identify such students and alert lecturers via a web application. 
Interestingly, Agnihotri et al \cite{3} showed that there is positive correlation between grades and login activity, but only up to certain level of activity, beyond which the effect diminishes. 

Lindrum \etal \cite{5} proposed early at-risk factor detection by measuring how well a final grade could be predicted by whether the student opened a course resource within a given time period.

Bovo \etal \cite{38} created an application to monitor students' progress during their course. For predicting progress, a number of different classifiers were considered, including logistic regression and naive bayes.
Input features to the classifiers included login frequency, date of last login, amount of time spent online, the number of lessons viewed, and the number of assignments completed. The authors used the mean grade obtained at the final exams as the target variable. Based on the outputs, the authors aimed to identify key predictors of final grades. In~\cite{47}, student grades were predicted by applying various Artificial Neural Networks to Moodle data for 250 students. ANNs with an incremental hidden layer algorithm turned produce the best results. Features which were used for the analysis included the number of examination sessions, mark, total accesses, percentage of resource views, total number of resources of each type viewed, and percentage of accesses per month. 

Jiang \etal \cite{28} analyzed data coming from Massive Open Online Courses (MOOCs), with a focus on student participation. Specifically, they attempted to predict final student performance based on a combination of students' Week 1 assignment performance and social interaction within the MOOC environment. Prediction was performed using  linear regression on a number of different features, including average quiz score, number of peer assessments, and their social network degree. 
\\*\\*
In the PASTA system \cite{4}, decision tree classifiers were used to predict whether a student would pass or fail their final exams. The predictions were based on information coming from three different sources: automatic marking system, discussion board, and assessment marks. Based on these predictions, timely preventive feedback was provided to students. Other systems have focused on providing course coordinators and administrators with insights into student progress. For instance, the DreamBox system visualizes the progress of the students in order to recommend the introduction of particular learning interventions to support students encountering difficulties~\cite{58}.

\section{Evaluation}
\label{sec:hypotheses}
Our main exercise was to test the three hypotheses outlined in the Introduction on our Moodle data. 
We find that the hypotheses are mostly supported by the data. 
The only exception is that there are a few scenarios where effort is \emph{negatively} correlated with grade. 

\subsection{Data}
The UCD Moodle data covered 360 courses, 2,194 assignments, and 71,077 assignment submissions. Student records were anonymized before any analysis was performed, and we considered only  data related to grades, deadlines, submission times, and general activity logs. For our experiments, we focused on a  subset of 60 assignments from courses run during first semester of the 2014/2015 academic year for which there was complete assignment submission information available.
\\
\\
\\
\subsection{Hypothesis 1}
\begin{figure}[!t]
\centering
\includegraphics[width=\linewidth]{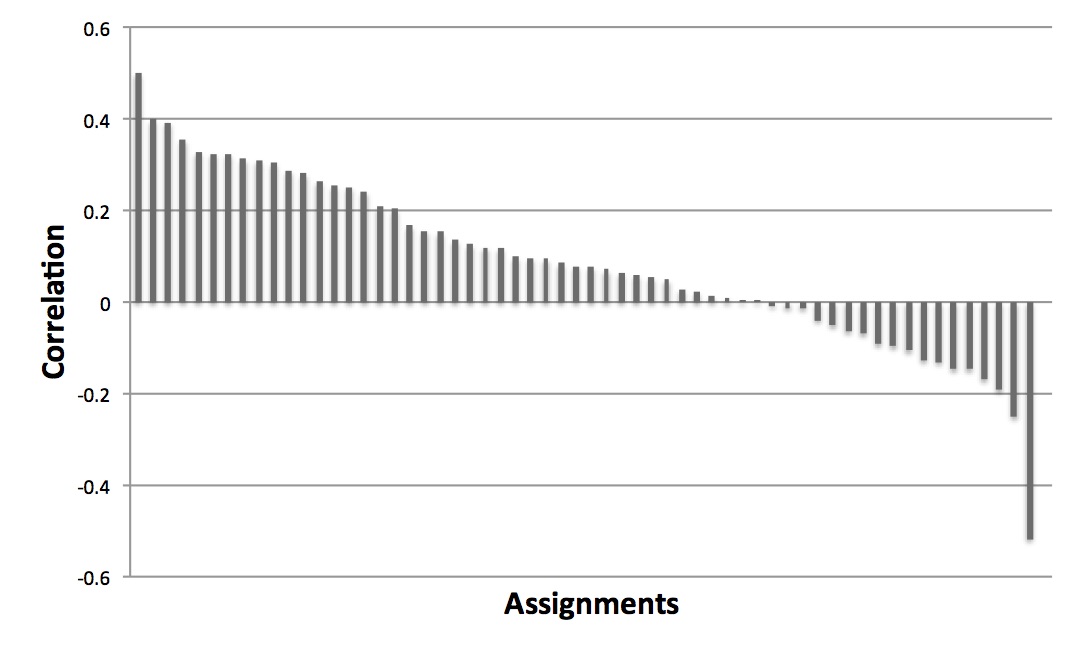}
\vskip -0.6em
\caption{Pearson correlations for all assignments between grades and time at which assignments were submitted, relative to assignment deadlines.}
\label{fig:correlations1}
\end{figure}

\begin{figure}[!t]
\centering
\includegraphics[width=\linewidth]{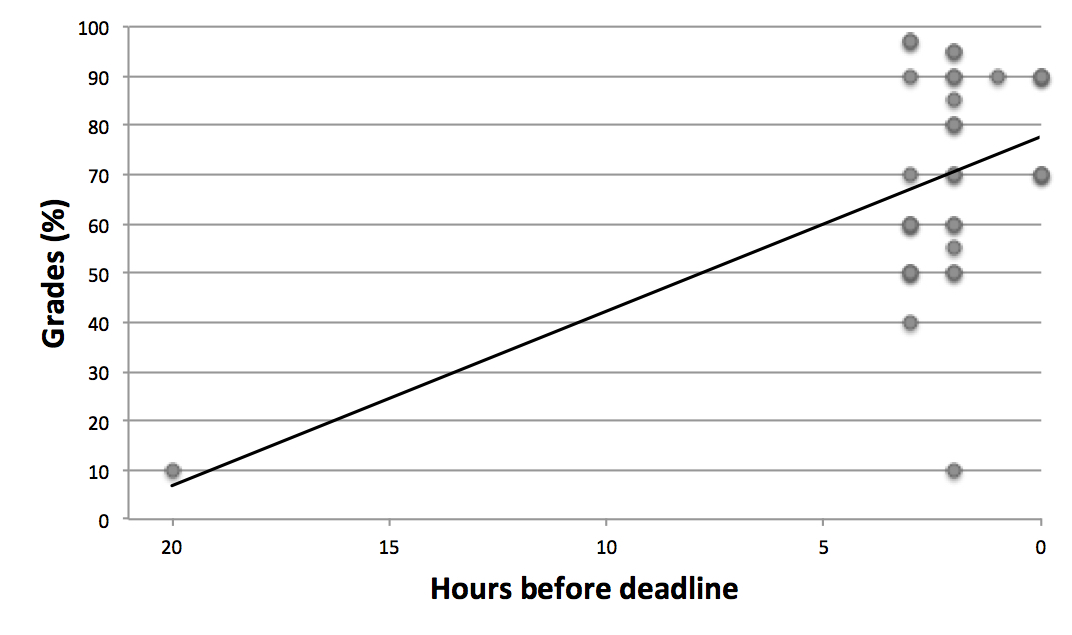}
\vskip 0.5em
\includegraphics[width=\linewidth]{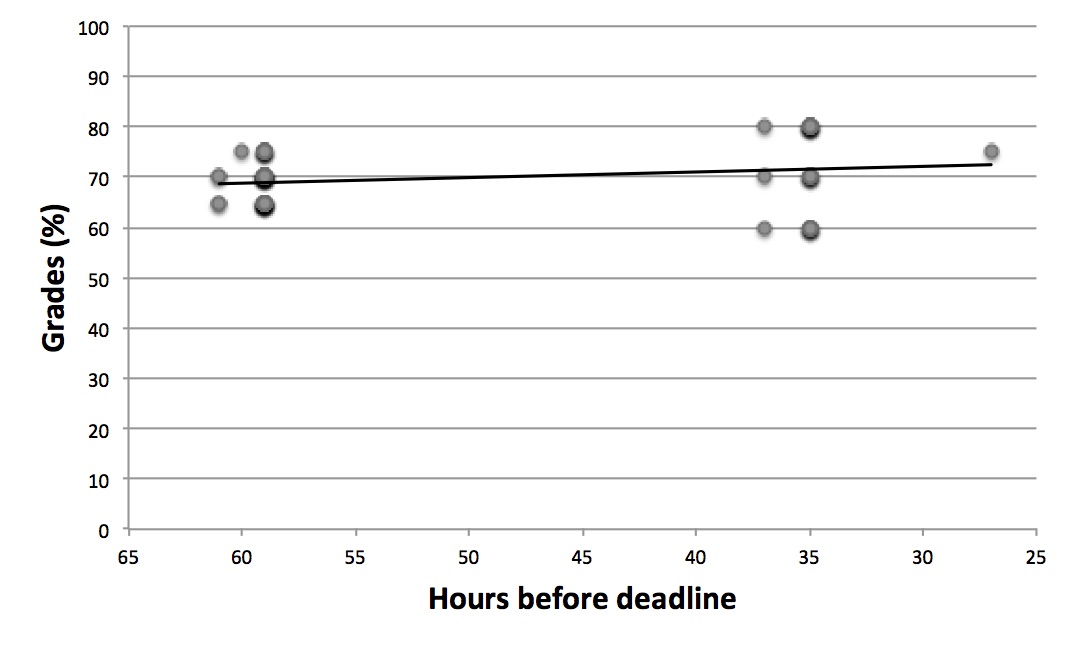}
\vskip -0.8em
\caption{Examples of courses with anomalies in relation to grades, when compared to number of hours before the assignment's submission deadline.}
\label{fig:anomaly}
\end{figure}

The expectation is that last minute submissions would not score well so the amount of time remaining between submission and the deadline should correlate with grade. From the analysis we excluded all  submissions after the deadline so that late penalties would not skew the results. We calculated correlation between grades and time remaining to submission for the 60 assignments. These correlations are sorted and plotted in Figure \ref{fig:correlations1}. Most of the assignments (42 out of 60) have positive correlation but there are some anomalies. Some reasons for these anomalies are; vaguely specified deadlines, deadline extensions, artefacts in marking schemes (see Figure \ref{fig:anomaly}), many submissions after deadline and no penalties for late submissions. 

The first graph in Figure \ref{fig:anomaly} shows that most of the submissions were very close to the deadline and an early submission with a very low grade skews the correlation. 
The second graph shows that the submissions are clustered at tutorial times (with 80 points stacked one on top of the other in a few places on the graph) with little variation between different grades.

\begin{figure}[!h]
\centering
\includegraphics[width=\linewidth]{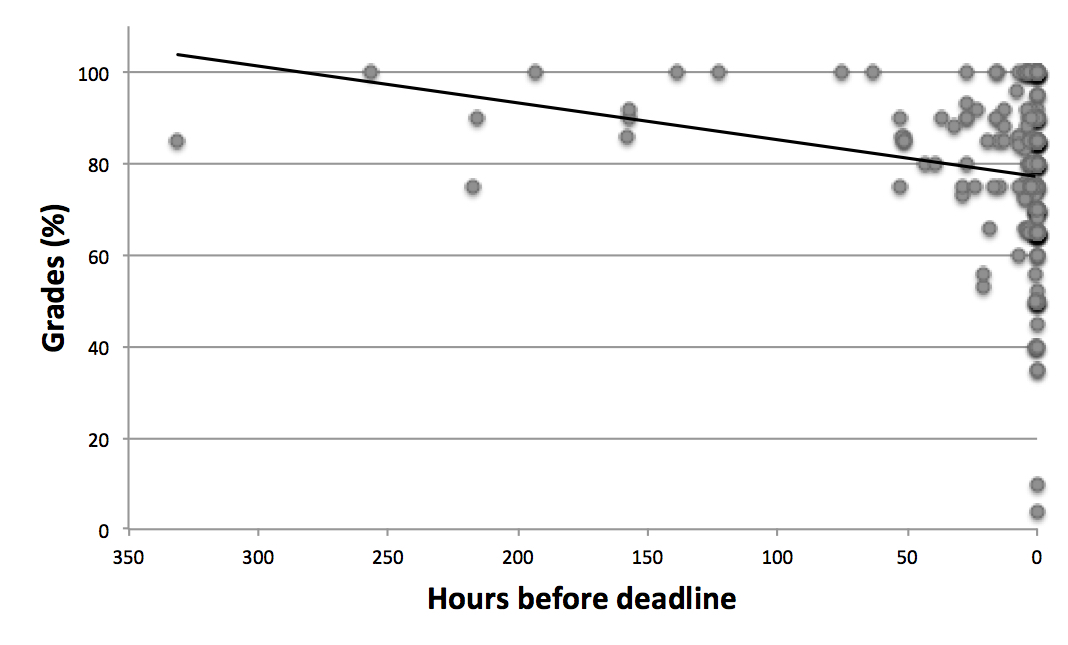}
\vskip -0.5em
\caption{Correlations for Level 3 assignments between grades and time at which assignments were submitted, relative to assignment deadlines (mean correlation = 0.33).}
\label{fig:level2}
\end{figure}

Most of the assignments with negative correlation are coming from Level 1 courses (1st year undergraduate), suggesting that few of the first year students have not developed good time management practices. 
We also find that Level 3 courses show the highest correlation between grades and submission time (see Figure \ref{fig:level2}). This suggests that most third year students are good at time management but the \emph{last minute} brigade are an obvious target for intervention. 

\subsection{Hypothesis 2}
The second hypothesis suggests that high levels of Moodle activity will correlate with a high grade. 
Overall activity plots confirm the findings reported in \cite{caseymining} about periods of time when students are more active on Moodle. We found that the first part of the week (Mon-Wed) and midday 
are most busy. We can infer from the results that high activity is caused by lectures and practical sessions. There are also spikes of activity in the evenings when students submit their assignments and work from home. The largest spikes in activity occur during examinations.

The main analysis conducted for hypothesis 2 is similar to that for the first hypothesis. However, due to the fact that each course includes a number of assignments and each might have different structure, time for submission, other factors influencing the submissions, we chose to do correlation analysis at an assignment level rather than a course level. 

Moodle logs many event types and it is to be expected that some events are not useful for this analysis. After analysis the correlations of different event subsets we settled on this subset \{\textsf{view assign, view course, view page, submit for grading assign, submission statement, accepted assign, submit assign, view confirm submit, assignment form assign}\}. This leaves a total of 241,401 events in the analysis down from an original set of 332,879.

\begin{figure}[!t]
\centering
\includegraphics[width=\linewidth]{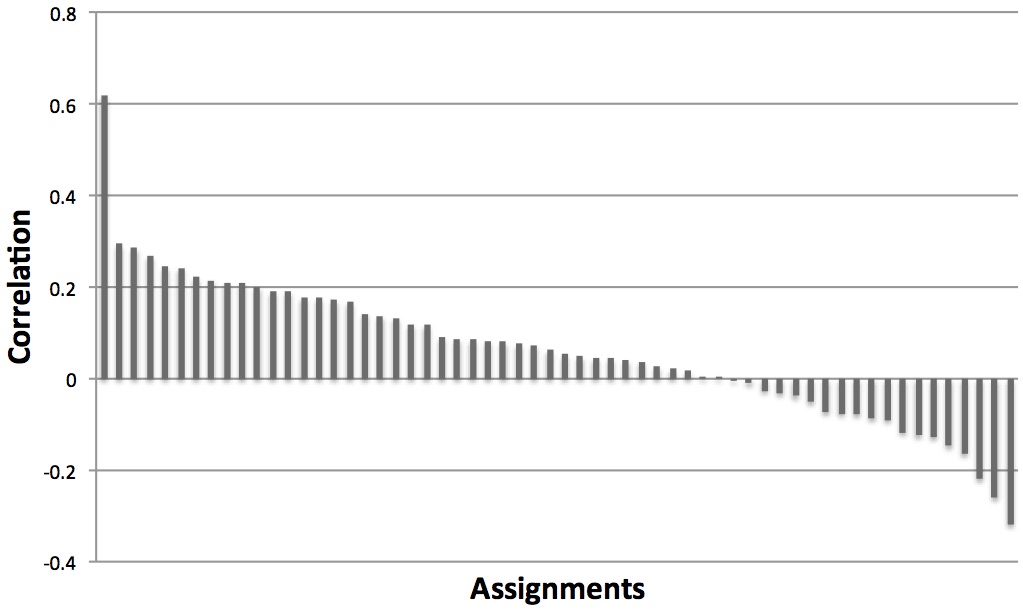}
\vskip -0.4em
\caption{Pearson correlations between grades and level of Moodle activity by students  for all assignments during the seven days before assignment submissions.}
\label{fig:correlations2}
\end{figure}

For each of the assignments we calculate the correlation between grade and activity as represented by Moodle events in the two weeks up to submission. These correlations are sorted and plotted in Figure \ref{fig:correlations2}. Most of the correlations (41 out of 60) are positive as expected, however the right hand side of the ranking shows some negative correlations. This is a bad sign as it suggests that effort does not always deliver a reward. These assignments are another target for intervention.



\subsection{Hypothesis 3}
Our final hypothesis suggests that evening activity should be more predictive 
of high grades than daytime activity. Testing this required that the activity data be 
divided into two intervals. To determine suitable time intervals, we examined the total number of activities per hour across all courses. The most significant spike of the daytime activity happened between 8am and 6pm, whereas evening activity rose between 6pm and midnight. 

\begin{table}[!t]
\centering
\caption{Summary of daytime and evening activity results, in terms of
(1) average count of activities per student during two weeks before
submission; (2) correlation between grades and number of activities.}
\vskip 0.7em
\label{tab:h3}
\begin{tabular}{|l|c|c|} \hline
\emph{Time} & \emph{Average count} & \emph{Grade-activity correlation}\\ \hline
Daytime & 41& 0.07\\ \hline
Evening& 15& 0.13\\
\hline\end{tabular}
\end{table}

Our analysis showed that evening activity did in fact show a higher correlation (Table \ref{tab:h3}), i.e. the correlation between evening activity levels and grade is stronger for daytime activity. This shows that the timing of activities is important in addition to the number of activities. This motivated the time-series analysis we report in the next section.

\section{Conclusions \& Future Work}
\label{sec:Conc}
Here we proposed a number of indicators of the success of students in achieving high grades in courses at undergraduate level. We used a large data set of Moodle log data originating from Computer Science courses to examine the extent to which these hypotheses were supported. While incomplete and noisy log data may have lead to correlations in our experiments that were not particularly strong, these correlations were still mostly positive, which suggests that the factors we have chosen are influencing the grades in the way we expect. 

The observations from our experiments further support the idea that students who are more active on Moodle and submit assignments earlier will achieve better results. As a next step, we plan to further explore the validity of the clusters classifying students into ``good'' and ``bad'' students on the assignment and course level. 

While we did observe outliers in our correlation analysis, the associated courses should be considered using a separate analysis to determine whether external factors are at play (e.g. continuous assessment rather than discrete assignments, lack of material provided on Moodle for a specific course). Finally, it is worth considering anomalous clusters in the context of activity outside that assignment or course.  In future research it would help in identification of the reason for the ``bad'' and ``good'' student performance, differences between modules and anomalies to change the way modules are delivered increasing successful learning.

\subsection*{Acknowledgments}
This publication has emanated from research conducted with the financial support of Science Foundation Ireland (SFI) under Grant Number SFI/12/RC/2289.


\bibliographystyle{plain}
\bibliography{moodle}


\balancecolumns
\end{document}